\documentclass[aip,jap,preprint,amsmath,amssymb]{revtex4-2}
\usepackage{graphicx}
\usepackage{booktabs}
\usepackage[version=4]{mhchem}
\usepackage{siunitx}
\usepackage{placeins}
\usepackage{xurl}
\usepackage[hidelinks]{hyperref}

\begin{document}

\title{Thermal Inverse Design: Tailoring the Transient Response of Resistive Micro-Heaters}

\author{Muzhang Huang}
\affiliation{Department of Aerospace Engineering, Korea Advanced Institute of Science and Technology (KAIST), Daejeon 34141, Republic of Korea}
\author{Junhe Zhang}
\affiliation{Department of Materials Science and Engineering, Massachusetts Institute of Technology, Cambridge, Massachusetts 02139, USA}
\author{Khoi Phuong Dao}
\affiliation{Department of Materials Science and Engineering, Massachusetts Institute of Technology, Cambridge, Massachusetts 02139, USA}
\author{Juejun Hu}
\affiliation{Department of Materials Science and Engineering, Massachusetts Institute of Technology, Cambridge, Massachusetts 02139, USA}
\author{Hyun Jung Kim}
\email{hyunjung.kim1@kaist.ac.kr}
\affiliation{Department of Aerospace Engineering, Korea Advanced Institute of Science and Technology (KAIST), Daejeon 34141, Republic of Korea}

\begin{abstract}
Microheaters underpin a broad range of applications that require localized and dynamic thermal control. Engineering their spatial and temporal temperature distributions is therefore essential, yet conventional heater design relies largely on iterative trial and error involving numerous full-scale thermal simulations. Here, we introduce an inverse design algorithm for pulsed heating, a common operating modality in thermal processing, sensing, switching, and material characterization. The approach uses a single lithographic step to pattern the doped regions of a silicon microheater and, in principle, enables on-demand peak temperature distributions under transient voltage excitation within one closed design loop. The algorithm also incorporates practical fabrication constraints, including finite lithographic resolution. We demonstrate its versatility through three representative designs. First, an optimized heater achieves more than a 20-fold reduction in temperature nonuniformity compared with a conventional uniformly doped heater. Second, inverse-designed doping effectively cancels the thermal gradient in a heater connected in series with a diode. Third, the method generates prescribed linear temperature gradients on demand. These results establish inverse-designed doping as a scalable and fabrication-compatible approach for tailoring transient temperature distributions in resistive microheaters.

\end{abstract}

\maketitle

\section{Introduction}
Resistive microheaters provide localized thermal actuation across electronics, microelectromechanical systems, microfluidics, and integrated photonics\cite{lovecchio2021largeAreaThinFilmHeater,lee2003designOptimisationMicrohotplate,iwaki2005designSimulationSOICMOSMicroheaters,rios2022ultraCompactNonvolatilePhaseShifter,zhong2021fastThermoOpticalModulators}. For many of these applications, the device performance is governed by the spatial temperature distribution across the functional region and, under pulsed operation, by its temporal evolution. Precise engineering of both the spatial and temporal temperature profiles is therefore essential for controlling thermal gradients, switching kinetics, material transformations, and device reliability.

Conventional microheater design relies largely on iterative trial and error, in which candidate geometries and operating conditions are repeatedly evaluated using full-scale thermal or electrothermal simulations\cite{lovecchio2021largeAreaThinFilmHeater,lee2003designOptimisationMicrohotplate,zhang2021electricallyReconfigurableMetasurface,yang2023designContinuouslyTunableVarifocal}. This process becomes computationally demanding as the number of geometric variables increases, particularly when temperature-dependent electrical and thermal properties must be included. Mathieu-Potvin and Gosselin optimized thin film heater topology and electrode potentials for prescribed temperature profiles\cite{mathieuPotvin2010optimalThinFilmTopologySpecifiedTemperature}, Tan et al. used nonlinear finite-element analysis, feedback, and Tikhonov regularization to determine the input currents of a microheater array for uniform and linear surface-temperature targets\cite{tan2017inverseHeatTransferMicroHeaterArray}, and Dao and Hu developed an inverse design scheme to achieve on-demand steady-state temperature profiles in doped Si microheaters\cite{dao2024thermalInverseDesignMicroHeaters}, which was experimentally validated to yield microheaters with considerably improved thermal uniformity\cite{tara2026inverse}. These studies established important approaches for controlling spatial temperature fields, but they primarily considered steady-state operation.

Transient temperature responses are technologically important in applications involving pulsed thermal processing, yet their spatial control has been explored far less extensively. During a finite voltage pulse, Joule heating and thermal diffusion evolve concurrently, such that a heater optimized for steady-state operation does not necessarily produce the desired temperature profile at a specified time. At elevated temperatures, changes in electrical conductivity, thermal conductivity, and heat capacity further couple the heater geometry, current distribution, and transient thermal response. Recent work has demonstrated manually engineered varying-width doped-silicon bridge heaters that tailor transient hotspots for repeatable multilevel phase-change operation\cite{sun2025microheaterHotspotEngineeringPCMPhotonics, vasquez2026electrothermal}. However, a general method for deriving a fabricable heater layout from a prescribed transient temperature field remains lacking.

The transient heater design problem can be formulated as an inverse problem: given a target spatial temperature field $T_{\mathrm{tar}}(\mathbf{r})$ at a prescribed time $t_{\mathrm{tar}}$, determine a realizable spatial actuation that reproduces it under a specified electrical pulse. The design variables may include local heat-source amplitudes, heater geometry, material or resistivity distributions, electrode arrangements, or electrical excitation. In the present work, the voltage pulse is fixed and the spatial actuation is encoded through variations in the doping pattern of silicon heaters. More generally, prescribed transient thermal fields may be implemented through topology, distributed electrical control, or geometric encoding of the heat source.

To solve the transient heater-design problem, three coupled challenges must be tackled. First, thermal diffusion produces strongly correlated responses from neighboring source regions, rendering fine-scale inversion ill-conditioned and susceptible to non-unique or rapidly oscillating power distributions. Second, the inferred power distribution must be translated into a continuous electrical layout that satisfies minimum-linewidth and spacing constraints. Third, temperature-dependent electrical and thermal properties create a nonlinear relationship among geometry, current density, power dissipation, and temperature, such that a design obtained from a linearized model generally requires device-level correction. Addressing these challenges requires an integrated framework that combines regularized field inversion, fabrication-constrained layout synthesis, and nonlinear electrothermal verification.

\begin{figure*}[htbp]
  \centering
\includegraphics[width=\textwidth]{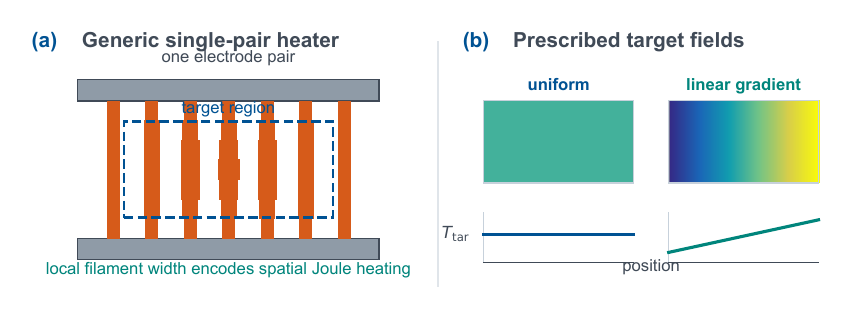}
  \caption{Resistive heater configuration and representative target temperature distributions. (a) The heater consists of multiple doped-silicon filaments connected in parallel between the same two electrodes. Varying the filament widths changes the local electrical resistance and redistributes Joule heating across the target region. (b) Representative prescribed temperature distributions at the target time $t_{\mathrm{tar}}$, including a uniform temperature and a linear temperature gradient.}
  \label{fig:problem}
\end{figure*}

To address these three challenges, we develop an inverse-design framework that integrates regularized field inversion, fabrication-constrained layout synthesis, and nonlinear electrothermal correction. The physical platform is a doped silicon microheater comprising multiple conductive filaments connected in parallel between a single pair of electrodes. Figure \ref{fig:problem} summarizes the heater configuration and representative target temperature fields. Each filament forms a continuous current path and is divided conceptually into segments whose widths can vary across the filament. Because the resistance of each segment depends on its width, spatial modulation of the filament widths redistributes Joule heating across the device. This relationship is coupled, since changing one segment affects both its local resistance and the current through the entire filament. Nevertheless, the complete heater can be fabricated using a single lithographic step that defines the binary doped and undoped regions, without requiring independently addressed heating elements.

The design procedure directly addresses the ill-conditioned nature of the inverse problem. A transient response matrix first relates groups of virtual heat sources to temperatures sampled across the target region at $t_{\mathrm{tar}}$. Physics-informed grouping reduces poorly distinguishable degrees of freedom, while constrained optimization determines the group powers required to approximate the prescribed temperature field. To ensure physical realizability, these powers are mapped to filament widths subject to minimum-linewidth and spacing constraints. Finally, the complete layout is evaluated using nonlinear transient electrothermal finite-element method (FEM) simulations to account for temperature-dependent material properties and geometry-induced current redistribution. The simulated temperature error is then used to update the power distribution and geometry to reduce deviations from the prescribed profile.

This closed-loop connection between transient field inversion, fabrication-constrained layout synthesis, and nonlinear device-level correction constitutes the central methodological advance. In the following, we demonstrate the approach by designing a uniform-temperature heater that reduces temperature nonuniformity by more than 20-fold relative to a uniformly doped heater operated at nearly identical electrical input power, compensating the thermal asymmetry introduced by a series-connected diode, and generating a prescribed linear temperature gradient.

\FloatBarrier

\section{Method}

\begin{figure*}[htbp]
  \centering
  \includegraphics[width=\textwidth]{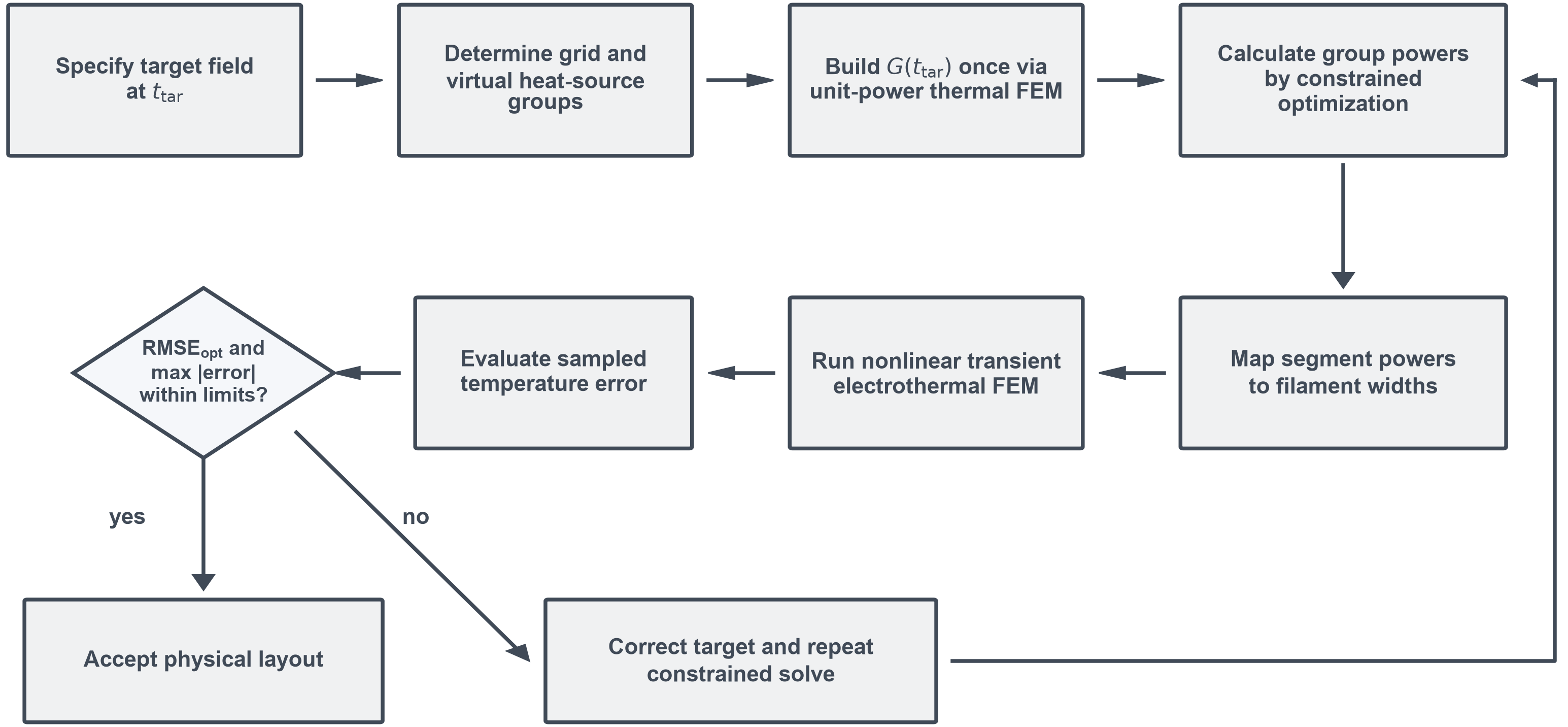}
  \caption{Overview of the inverse-design procedure. The prescribed temperature distribution, target time, voltage pulse, and fabrication constraints are first specified. The heater is divided into design cells, which are assigned to virtual heat-source groups according to the heater geometry and target distribution. Transient thermal simulations are used to construct a response matrix, and constrained optimization determines the required power for each group. The optimized powers are converted into filament widths and evaluated using nonlinear transient electrothermal FEM. If the simulated temperature does not satisfy the error criteria, the power distribution and filament widths are updated and evaluated again. $\mathrm{RMSE}_{\mathrm{opt}}$ denotes the root-mean-square error of the boundary-averaged cell temperatures relative to the target.}
  \label{fig:method_workflow}
\end{figure*}

Figure \ref{fig:method_workflow} summarizes the inverse-design procedure. The design inputs are a target temperature field $T_{\mathrm{tar}}(\mathbf{r})$, a target time $t_{\mathrm{tar}}$, a prescribed voltage-pulse waveform, and the geometric and fabrication constraints of the heater. The procedure comprises four steps. First, the heater domain is partitioned into design cells and cells with equivalent physical roles are assigned to virtual heat-source groups. Second, transient simulations of these groups are used to construct a response matrix relating their power densities to the temperature field at $t_{\mathrm{tar}}$. Third, the optimized cell powers are converted into the widths of continuous doped silicon filaments connected between a single pair of electrodes. Finally, the resulting physical layout is evaluated using nonlinear transient electrothermal FEM, and the simulated error is used to correct the power distribution and geometry. This process is repeated until both the thermal-error and fabrication criteria are satisfied.

\subsection{Design-cell definition and size selection}

The patterned heater domain, denoted by $\Omega_{\mathrm{H}}$, is partitioned into rectangular design cells $\Omega_{ij}$. Index $i$ specifies the position transverse to the current direction and therefore identifies a conducting filament, while index $j$ identifies successive segments along that filament. Each cell has an in-plane width $W_i$, a length $L_j$ along the current direction, and the silicon-device-layer thickness $h$. During the inverse calculation, the heat-generation density within each cell is treated as spatially uniform. After power-to-geometry mapping, the same cell contains one segment of a continuous doped-silicon filament.

The design cells are distinct from the finite elements used in the forward simulations. A design cell defines the spatial resolution of the inverse-design variables and the physical pitch of the heater filaments. Each design cell is discretized by multiple finite elements, and the FEM mesh is refined independently until further refinement produces negligible changes in the calculated temperature field.

Cell size is selected by balancing temperature uniformity against fabrication constraints. Within the inverse model, each cell is represented by a single heat-generation value. The cell must therefore be sufficiently small that the temperature variation within its footprint remains acceptable. In the physical heater, however, Joule heating is concentrated in a doped filament occupying only part of the cell width. This discrete arrangement produces a local temperature maximum above each filament and a minimum between neighboring filaments. Increasing the cell size increases the spacing between these localized heat sources and consequently amplifies the fine-scale temperature oscillations, as we will show later. A smaller cell pitch reduces these oscillations and more closely approximates a continuous spatial heat-source distribution.

Reducing the cell size, however, also decreases the power assigned to each cell and generally produces narrower filament segments after the power-to-linewidth conversion. It can therefore reduce both the minimum linewidth and the available line-to-line spacing. The cell pitch is consequently chosen as the smallest value that can be implemented while maintaining the required fabrication margins. In practice, candidate cell sizes are evaluated by first performing the inverse power optimization and linewidth mapping, followed by nonlinear electrothermal simulation of the resulting layout. An acceptable cell size must simultaneously limit the within-cell temperature variation and satisfy the minimum linewidth and spacing rules across the complete heater.

Two central-cell sizes are considered in this work to illustrate this trade-off. The square heater uses $1~\mu\mathrm{m}\times1~\mu\mathrm{m}$ cells, which provide finer spatial sampling and smaller local temperature oscillations. A second rectangular heater uses $2~\mu\mathrm{m}\times2~\mu\mathrm{m}$ cells, which yield larger linewidth, spacing margins, and hence fabrication tolerance at the expense of increased fine-scale temperature variation. The same $2~\mu\mathrm{m}$ cell size is also used for the linear-gradient design.

A $4~\mu\mathrm{m}$-wide peripheral compensation cell is added along each side of the central cell array. These peripheral cells are included in the inverse model because they provide the additional heat generation needed to compensate for enhanced heat loss near the heater boundaries. They are not included in the target region over which the temperature error is evaluated. The resulting square design comprises a central $42\times42$ array of $1~\mu\mathrm{m}$ cells surrounded by one compensation cell on each side, giving a $44\times44$ grid over a $50~\mu\mathrm{m}\times50~\mu\mathrm{m}$ heater. The rectangular designs comprise a central $31\times21$ array of $2~\mu\mathrm{m}$ cells surrounded by $4~\mu\mathrm{m}$ compensation cells, giving a $33\times23$ grid over a $70~\mu\mathrm{m}\times50~\mu\mathrm{m}$ heater.

For each design cell in the target region, temperatures are sampled at the mid-plane of the silicon device layer at the cell center and at the two cell boundaries transverse to the current direction. The scalar used in the response-matrix optimization is the mean of the two boundary temperatures; the center value is retained as a diagnostic of the within-cell temperature variation. Only the two transverse boundaries are used because the filaments are oriented along the current direction, so the filament-to-gap temperature oscillation occurs primarily across the filaments; the other two cell boundaries lie along the filament and do not sample this pitch-induced variation. Thus, $N$ is the number of target-region cells: $N=42\times42=1764$ for the $1~\mu\mathrm{m}$ square design and $N=31\times21=651$ for each $2~\mu\mathrm{m}$ rectangular design. Index $n$ below denotes one target-region cell, and $T_{\mathrm{tar}}(\mathbf{r}_n)$ is its prescribed temperature. The target-temperature vector is

\begin{equation}
    \mathbf{T}_{\mathrm{tar}}
    =
    \begin{bmatrix}
        T_{\mathrm{tar}}(\mathbf{r}_1) &
        T_{\mathrm{tar}}(\mathbf{r}_2) &
        \cdots &
        T_{\mathrm{tar}}(\mathbf{r}_N)
    \end{bmatrix}^{\mathrm{T}}.
    \label{eq:target_vector}
\end{equation}

The same boundary-averaged cell temperatures are used to construct the response matrix, perform the inverse optimization, and evaluate convergence. Temperatures throughout and between the cells are subsequently assessed using the complete nonlinear electrothermal model, thereby capturing the fine-scale oscillations produced by the discrete filaments.

\subsection{Virtual heat-source grouping and response matrix}

Assigning an independent power to every cell would create many nearly indistinguishable design variables because thermal diffusion strongly couples neighboring cells. We therefore group cells that have equivalent roles under the symmetry of the heater and the target field. Let $g(i,j)\in\{1,\ldots,M\}$ identify the virtual group containing cell $\Omega_{ij}$, where $M$ is the number of independently optimized groups. The binary mask of group $m$ is defined as

\begin{equation}
    B_m(\mathbf{r})
    =
    \begin{cases}
        1, & \mathbf{r}\in\Omega_{ij}\ \mathrm{and}\ g(i,j)=m,\\
        0, & \mathrm{otherwise}.
    \end{cases}
    \label{eq:group_mask}
\end{equation}

\begin{figure*}[htbp]
  \centering
  \includegraphics[width=\textwidth]{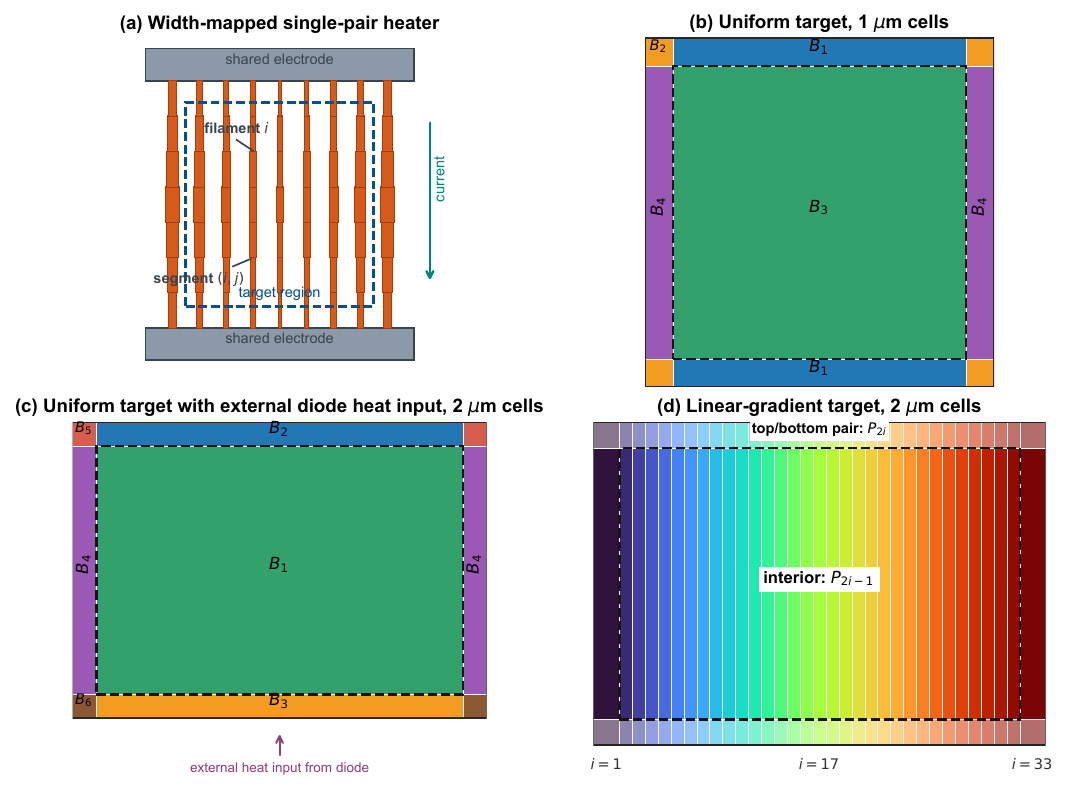}
\caption{Physical heater representation and architecture-dependent virtual heat-source grouping. (a) Continuous doped-silicon filaments share one pair of electrodes, while their segment widths vary both along and transverse to the current direction. (b) The square uniform-temperature design uses four groups that preserve both in-plane symmetries. (c) The rectangular uniform-temperature design in the one-diode--one-resistor (1D1R) configuration uses six groups; separate upper and lower edge and corner variables allow compensation of the one-sided external heat input from the series diode. (d) For the linear-gradient target, each column $i$ has an independently optimized interior variable $P_{2i-1}$ and a variable $P_{2i}$ shared by its top and bottom compensation segments. Dashed outlines mark the target regions. Colors indicate group membership, not power or temperature.}
  \label{fig:heater_layout}
\end{figure*}

The virtual heat-source field used to construct the response matrix is

\begin{equation}
    q(\mathbf{r},t)
    =
    f(t)
    \sum_{m=1}^{M}
    B_m(\mathbf{r})p_m,
    \label{eq:virtual_heat_source}
\end{equation}

\noindent where $q(\mathbf{r},t)$ is the volumetric heat-generation rate, $f(t)$ is the dimensionless pulse envelope, and $p_m$ is the heat-generation amplitude assigned to group $m$. The masks $B_m(\mathbf{r})$ describe where heat is generated, while $p_m$ determines its magnitude. These virtual sources are numerical constructs used only during inverse design. In the final device, their optimized powers are encoded in the widths of a continuous resistive heater. Figure~\ref{fig:heater_layout} shows the four-group square design, the six-group rectangular design used with one-sided diode-associated heating, and the 66-variable grouping used for the linear-gradient target.

For a fixed pulse waveform, a reference simulation is performed for each group. In simulation $m$, a reference power density $p_{\mathrm{ref}}$ is applied only to mask $B_m(\mathbf{r})$. The response-matrix element

\begin{equation}
    G_{nm}(t_{\mathrm{tar}})
    =
    \frac{
        T_n^{(m)}(t_{\mathrm{tar}})-T_{\mathrm{amb}}
    }{
        p_{\mathrm{ref}}
    }
    \label{eq:response_matrix_element}
\end{equation}

\noindent gives the boundary-averaged temperature rise for target-region cell $n$ per unit power density applied to group $m$. Here, $T_n^{(m)}$ is the mean of the two transverse-boundary temperatures for cell $n$ in reference simulation $m$, and $T_{\mathrm{amb}}$ is the ambient temperature. Each column of $\mathbf{G}$ is therefore the transient thermal influence function of one virtual source group. It includes lateral and vertical heat diffusion, heat loss through the substrate and exterior boundaries, and thermal coupling among different regions of the heater.

With material properties held fixed during construction of $\mathbf{G}$, the individual responses can be superposed. The predicted temperature vector at $t_{\mathrm{tar}}$ is

\begin{equation}
    \mathbf{T}_{\mathrm{pred}}(t_{\mathrm{tar}})
    =
    T_{\mathrm{amb}}\mathbf{1}_N
    +
    \mathbf{G}(t_{\mathrm{tar}})\mathbf{p},
    \label{eq:linear_response}
\end{equation}

\noindent where $\mathbf{1}_N$ is an $N$-component vector of ones and

\begin{equation}
    \mathbf{p}
    =
    \begin{bmatrix}
        p_1 & p_2 & \cdots & p_M
    \end{bmatrix}^{\mathrm{T}}
\end{equation}

\noindent contains the group power densities. Equation~\eqref{eq:linear_response} is a reduced linear model evaluated only at the prescribed time. It does not assume that the target temperature profile is maintained throughout the pulse.

The initial group powers are determined from

\begin{equation}
    \mathbf{p}^{(0)}
    =
    \operatorname*{arg\,min}_{\mathbf{p}\in\mathcal{C}}
    \left[
        \left\|
            \mathbf{G}\mathbf{p}
            -
            \left(
                \mathbf{T}_{\mathrm{tar}}
                -
                T_{\mathrm{amb}}\mathbf{1}_N
            \right)
        \right\|_2^2
        +
        \lambda
        \left\|
            \mathbf{L}\mathbf{p}
        \right\|_2^2
    \right].
    \label{eq:initial_optimization}
\end{equation}

The first term measures the mismatch between the predicted and prescribed temperature rises. The feasible set $\mathcal{C}$ enforces non-negative heat generation and the specified power-contrast bounds. Cell grouping is incorporated directly through the reduced vector $\mathbf{p}$. For the uniform-temperature designs, symmetry-based grouping provides sufficient regularization and $\lambda=0$. For the linear-gradient design, $\mathbf{L}$ is a second-difference operator applied along the gradient direction, and $\lambda>0$ suppresses rapidly alternating power values between adjacent columns without suppressing the prescribed linear trend. Thus, the second term penalizes spatial variations that cannot be resolved thermally or implemented robustly in the physical heater.

\subsection{Conversion from optimized power to filament width}

After optimization, each group power density is expanded to the cells belonging to that group. The power assigned to cell $(i,j)$ is obtained by integrating the virtual heat source over the cell volume:

\begin{equation}
    P_{ij}
    =
    \int_{\Omega_{ij}}
    q(\mathbf{r},t_{\mathrm{on}})\,\mathrm{d}V,
    \label{eq:cell_power}
\end{equation}

\noindent where $t_{\mathrm{on}}$ denotes a time during the constant-amplitude portion of the pulse. For a uniform volumetric power density within a cell, Eq.~\eqref{eq:cell_power} reduces to $P_{ij}=p_{g(i,j)}W_iL_jh$.

Each fixed-$i$ column is mapped to a continuous doped silicon filament connecting the two electrodes. All segments within filament $i$ are connected in series, while the different filaments are connected in parallel and experience the same applied voltage $V$. The total power assigned to filament $i$ is

\begin{equation}
    P_i
    =
    \sum_j P_{ij}.
    \label{eq:filament_power}
\end{equation}

Because the electrical power dissipated by the complete filament is $P_i=VI_i$, its current is

\begin{equation}
    I_i
    =
    \frac{P_i}{V},
    \label{eq:filament_current}
\end{equation}

\noindent where $I_i$ is the current flowing through every series-connected segment of filament $i$. The resistance required for segment $(i,j)$ to dissipate power $P_{ij}$ is then obtained from Joule's law:

\begin{equation}
    R_{ij}
    =
    \frac{P_{ij}}{I_i^2}.
    \label{eq:segment_resistance}
\end{equation}

Approximating the segment as a rectangular doped silicon conductor gives

\begin{equation}
    R_{ij}
    =
    \frac{\rho_{\mathrm{e}}L_j}{h w_{ij}},
    \label{eq:rectangular_resistance}
\end{equation}

\noindent where $\rho_{\mathrm{e}}$ is the electrical resistivity used for the initial geometric mapping and $w_{ij}$ is the required segment width. Combining Eqs.~\eqref{eq:segment_resistance} and \eqref{eq:rectangular_resistance} yields

\begin{equation}
    w_{ij}
    =
    \frac{\rho_{\mathrm{e}}L_j}{hR_{ij}}
    =
    \frac{\rho_{\mathrm{e}}L_jI_i^2}{hP_{ij}}.
    \label{eq:linewidth_mapping}
\end{equation}

Equations~\eqref{eq:filament_power}--\eqref{eq:linewidth_mapping} provide the physical connection between the inverse thermal solution and the lithographic heater pattern. The optimized power in a segment determines its required resistance, and the resistance is encoded through the segment width. Because changing one segment also changes the total resistance and current of its filament, the complete set of segment widths must be evaluated as one electrically coupled layout rather than as independent local heaters.

For adjacent filaments centered in cells of widths $W_i$ and $W_{i+1}$, the edge-to-edge spacing at row $j$ is

\begin{equation}
    s_{i,i+1,j}
    =
    \frac{W_i-w_{ij}}{2}
    +
    \frac{W_{i+1}-w_{i+1,j}}{2}.
    \label{eq:filament_spacing}
\end{equation}

A mapped design is geometrically admissible only when

\begin{equation}
    w_{ij}\geq w_{\min},
    \qquad
    s_{i,i+1,j}\geq s_{\min}
    \label{eq:fabrication_constraints}
\end{equation}

\noindent throughout the complete heater, including the peripheral compensation cells. Here, $w_{\min}$ and $s_{\min}$ are the minimum linewidth and line-to-line spacing allowed by the fabrication process. Figure~\ref{fig:heater_layout}(a) illustrates the resulting layout, in which the widths vary both among adjacent filaments and among successive segments of the same filament.

\subsection{Nonlinear electrothermal verification and correction}

The response matrix provides an efficient initial design but does not include the complete electrical feedback of the linewidth-mapped heater. The final physical model solves the coupled electrical and transient heat equations:

\begin{align}
    \nabla\cdot
    \left[
        \sigma(T)\nabla\phi
    \right]
    &=
    0,
    \label{eq:electrical_governing}
    \\
    q_{\mathrm{J}}
    &=
    \sigma(T)
    \left|
        \nabla\phi
    \right|^2,
    \label{eq:joule_heating}
    \\
    \rho_{\mathrm{m}}C_p(T)
    \frac{\partial T}{\partial t}
    &=
    \nabla\cdot
    \left[
        k(T)\nabla T
    \right]
    +
    q_{\mathrm{J}}.
    \label{eq:thermal_governing}
\end{align}

Here, $\phi$ is the electric potential, $\sigma(T)$ is the temperature-dependent electrical conductivity, $q_{\mathrm{J}}$ is the local Joule-heating density, $\rho_{\mathrm{m}}$ is the mass density, $C_p(T)$ is the specific heat capacity, and $k(T)$ is the thermal conductivity. Equation~\eqref{eq:electrical_governing} determines the current redistribution produced by the spatially varying filament widths. Equation~\eqref{eq:joule_heating} converts the resulting electric field into a volumetric heat source, and Eq.~\eqref{eq:thermal_governing} describes the accumulation and diffusion of heat during the pulse.

The forward model uses a silicon-on-insulator stack comprising a $0.22~\mu\mathrm{m}$ silicon device layer, a $2~\mu\mathrm{m}$ buried oxide layer, and a $20~\mu\mathrm{m}$ silicon substrate. The bottom of the substrate is maintained at $T_{\mathrm{amb}}=293.15~\mathrm{K}$, and a convective heat-flux condition is applied to the top surface. The applied waveform is a rectangular step pulse: the voltage is switched from zero to its stated amplitude at $t=0$ and held constant through the target time $t_{\mathrm{tar}}=13~\mu\mathrm{s}$. The silicon heat capacity is obtained from the NIST--JANAF Shomate correlation\cite{chase1998nistJanafThermochemicalTables}. Its thermal conductivity is represented by

\begin{equation}
    k_{\mathrm{Si}}(T)
    =
    k_{293}
    \frac{293.15~\mathrm{K}}{T},
    \label{eq:silicon_thermal_conductivity}
\end{equation}

\noindent where $k_{293}$ is the silicon thermal conductivity at $293.15~\mathrm{K}$ \cite{shanks1963thermalConductivitySilicon,glassbrenner1964thermalConductivitySiliconGermanium}. The doped-silicon electrical conductivity $\sigma(T)$ is interpolated from the temperature-dependent measurements reported by Popescu et al. \cite{popescu2023openSourceMultifunctionalTestingPlatform}. Further material parameters and boundary conditions are provided in Sec.~S1 of the Supplementary Material.

At correction iteration $k$, the FEM simulated temperature vector is compared with the prescribed target:

\begin{equation}
    \mathbf{e}^{(k)}
    =
    \mathbf{T}_{\mathrm{FEM}}^{(k)}
    -
    \mathbf{T}_{\mathrm{tar}},
    \label{eq:nonlinear_error}
\end{equation}

\noindent where $\mathbf{T}_{\mathrm{FEM}}^{(k)}$ contains the boundary-averaged temperatures of the same $N$ target-region cells used to construct the response matrix. A positive component of $\mathbf{e}^{(k)}$ indicates that the physical heater is too hot in the corresponding cell, whereas a negative component indicates insufficient heating. The corrected target supplied to the linear model is

\begin{equation}
    \mathbf{T}_{\mathrm{corr}}^{(k)}
    =
    T_{\mathrm{amb}}\mathbf{1}_N
    +
    \mathbf{G}\mathbf{p}^{(k)}
    -
    \gamma_k\mathbf{e}^{(k)},
    \label{eq:corrected_target}
\end{equation}

\noindent where $\gamma_k$ is the correction gain. The next group-power vector is found from

\begin{equation}
    \mathbf{p}^{(k+1)}
    =
    \operatorname*{arg\,min}_{\mathbf{p}\in\mathcal{C}}
    \left[
        \left\|
            T_{\mathrm{amb}}\mathbf{1}_N
            +
            \mathbf{G}\mathbf{p}
            -
            \mathbf{T}_{\mathrm{corr}}^{(k)}
        \right\|_2^2
        +
        \lambda
        \left\|
            \mathbf{L}\mathbf{p}
        \right\|_2^2
    \right].
    \label{eq:nonlinear_correction}
\end{equation}

The updated powers are converted into a new linewidth map using Eqs.~\eqref{eq:filament_power}--\eqref{eq:linewidth_mapping}, after which the nonlinear electrothermal model is solved again. If an update increases the maximum absolute temperature error, $\gamma_k$ is reduced and the update is recomputed from the preceding accepted design.

Convergence is assessed using the optimization-sample RMSE and maximum absolute error:

\begin{align}
    \mathrm{RMSE}_{\mathrm{opt}}^{(k)}
    &=
    \sqrt{
        \frac{1}{N}
        \sum_{n=1}^{N}
        \left(
            e_n^{(k)}
        \right)^2
    },
    \label{eq:rmse}
    \\
    e_{\max}^{(k)}
    &=
    \max_n
    \left|
        e_n^{(k)}
    \right|.
    \label{eq:max_error}
\end{align}

Convergence of the retained designs is assessed using the boundary-averaged cell temperatures, with common tolerances of $\mathrm{RMSE}_{\mathrm{opt}}^{(k)}\leq2~\mathrm{K}$ and $e_{\max}^{(k)}\leq5~\mathrm{K}$. If further acceptable updates cannot be found, correction stops and the last accepted design is retained even if these tolerances are not met. The linewidth and spacing requirements in Eq.~\eqref{eq:fabrication_constraints} are enforced throughout. This final correction step accounts for temperature-dependent material properties, current redistribution, and geometric effects that are absent from the linear response matrix.

For the two-dimensional temperature fields reported in the Results, we define
\begin{equation}
    \mathrm{RMSE}_{\mathrm{field}}
    =
    \sqrt{\frac{1}{N_{\mathrm{field}}}
    \sum_{\ell=1}^{N_{\mathrm{field}}}
    \left[T(x_\ell,y_\ell)-T_{\mathrm{tar}}(x_\ell,y_\ell)\right]^2}.
    \label{eq:rmse_field}
\end{equation}
Here, $N_{\mathrm{field}}$ is the number of uniformly spaced spatial sampling points retained within the target region at the device-layer mid-plane. The sampling grid divides the full heater extent into 700 equal intervals along each axis for the square heaters, and 800 along $x$ and 600 along $y$ for the rectangular heaters. Only sampling points within the target region are retained for calculating $N_{\mathrm{field}}$ and $\mathrm{RMSE}_{\mathrm{field}}$. The FEM temperature at $t_{\mathrm{tar}}$ is interpolated onto this grid, independently of the FEM mesh and design-cell discretization. All points have equal weight and cover both filament interiors and gaps. The maximum positive and negative field errors use the same pointwise differences relative to the target.

\FloatBarrier
\section{Results}

\subsection{Heaters for Producing a Uniform Temperature Distribution}
\label{subsec:1um}

The first case considers a $50~\mu\mathrm{m}\times50~\mu\mathrm{m}$ heater containing a central $42~\mu\mathrm{m}\times42~\mu\mathrm{m}$ target region and a $4~\mu\mathrm{m}$-wide peripheral compensation region along each side. The inverse-designed heater is compared with a reference heater comprising a continuous silicon region with uniform doping. The two heaters have the same layer stack, overall dimensions, target region, and target time of $13~\mu\mathrm{s}$. The voltage applied to the reference heater is adjusted so that the two heaters operate at nearly identical electrical input powers of approximately $1.23~\mathrm{W}$.

\begin{figure*}[htbp]
    \centering
    \includegraphics[width=\linewidth,height=0.66\textheight,keepaspectratio]{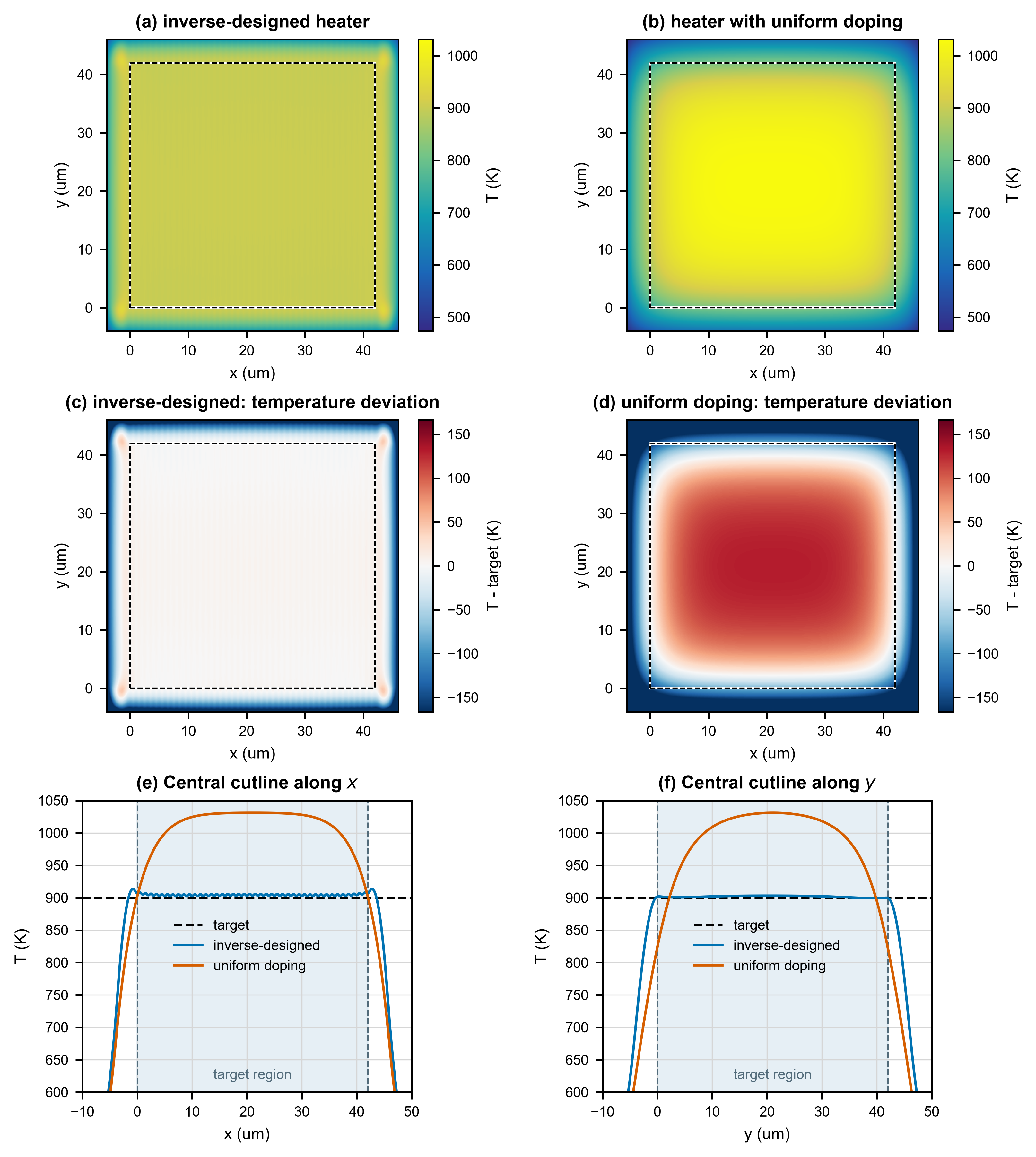}
    \caption{Comparison of the inverse-designed heater and the reference heater with uniform doping under nearly identical electrical input power. The temperature fields are evaluated at $t=13~\mu\mathrm{s}$ at the mid-plane of the silicon device layer. (a,b) Temperature distributions produced by the inverse-designed and reference heaters. (c,d) Temperature deviations from the prescribed $900~\mathrm{K}$ target. (e,f) Temperature profiles along the $x$ and $y$ directions through the center of the target region, respectively. The dashed rectangles in (a--d) and shaded intervals in (e,f) indicate the target region. The inverse-designed heater suppresses the large center-to-edge temperature variation observed in the reference heater. The remaining small periodic variation originates from the discrete doped silicon filaments.}
    \label{fig:one_um_plain_comparison}
\end{figure*}

Figure \ref{fig:one_um_plain_comparison} shows the temperature fields evaluated at the mid-plane of the $220~\mathrm{nm}$ silicon device layer. Under nearly identical input power, the reference heater with uniform doping exhibits a pronounced center-to-edge temperature variation. In contrast, the inverse-designed heater redistributes Joule heating from the center toward the perimeter and produces a substantially more uniform temperature field. Within the target region, the $\mathrm{RMSE}_{\mathrm{field}}$ is $3.3~\mathrm{K}$ for the inverse-designed heater, compared with $90.1~\mathrm{K}$ for the reference heater. The maximum positive and negative deviations from the $900~\mathrm{K}$ target are $+8.1~\mathrm{K}$ and $-4.4~\mathrm{K}$ for the inverse-designed heater, compared with $+131.0~\mathrm{K}$ and $-166.2~\mathrm{K}$ for the reference heater. The inverse design therefore reduces the maximum absolute temperature error by more than a factor of 20.

Small periodic variations remain in the temperature field of the inverse-designed heater because Joule heating is concentrated within discrete doped silicon filaments. These variations decrease as heat spreads laterally during the pulse.

The smallest filament-segment width in the final inverse-designed heater is $280.6~\mathrm{nm}$, and the smallest edge-to-edge spacing between neighboring filaments is $502.3~\mathrm{nm}$. These values are evaluated over the complete heater, including the peripheral compensation cells. Both dimensions can be increased by using larger design cells. In the next section, we demonstrate a heater with $2~\mu\mathrm{m}$ cells, which provides greater fabrication tolerance at the expense of reduced temperature uniformity arising from larger within-cell temperature oscillations.

\FloatBarrier
\subsection{Uniform-Temperature Heater Design in a 1D1R Addressing Architecture}
\label{sec:rectangular_heater}

We next apply the inverse-design method to a heater integrated into a one-diode--one-resistor (1D1R) addressing architecture. This configuration is used in two-dimensional crossbar arrays, where each pixel contains a silicon p--intrinsic--n (PIN) diode connected in series with a resistive heater\cite{popescu2026twoDimensionalAddressableMetasurface}. The diode serves as a selector that suppresses sneak-path current through unaddressed pixels. During addressed operation, the same current passes through the diode and the resistive heater.

The heater occupies an area of $70~\mu\mathrm{m}\times50~\mu\mathrm{m}$. A central $31\times21$ array of $2~\mu\mathrm{m}\times2~\mu\mathrm{m}$ cells defines the $62~\mu\mathrm{m}\times42~\mu\mathrm{m}$ target region. One $4~\mu\mathrm{m}$-wide peripheral compensation cell is added along each side, giving a total design grid of $33\times23$ cells. Compared with the $1~\mu\mathrm{m}$ cells used in Sec. \ref{subsec:1um}, the larger cells provide greater minimum linewidth and spacing, although they also increase the local temperature variation between the center of each filament and the space between neighboring filaments.

\begin{figure*}[htbp]
    \centering
    \includegraphics[width=\linewidth]{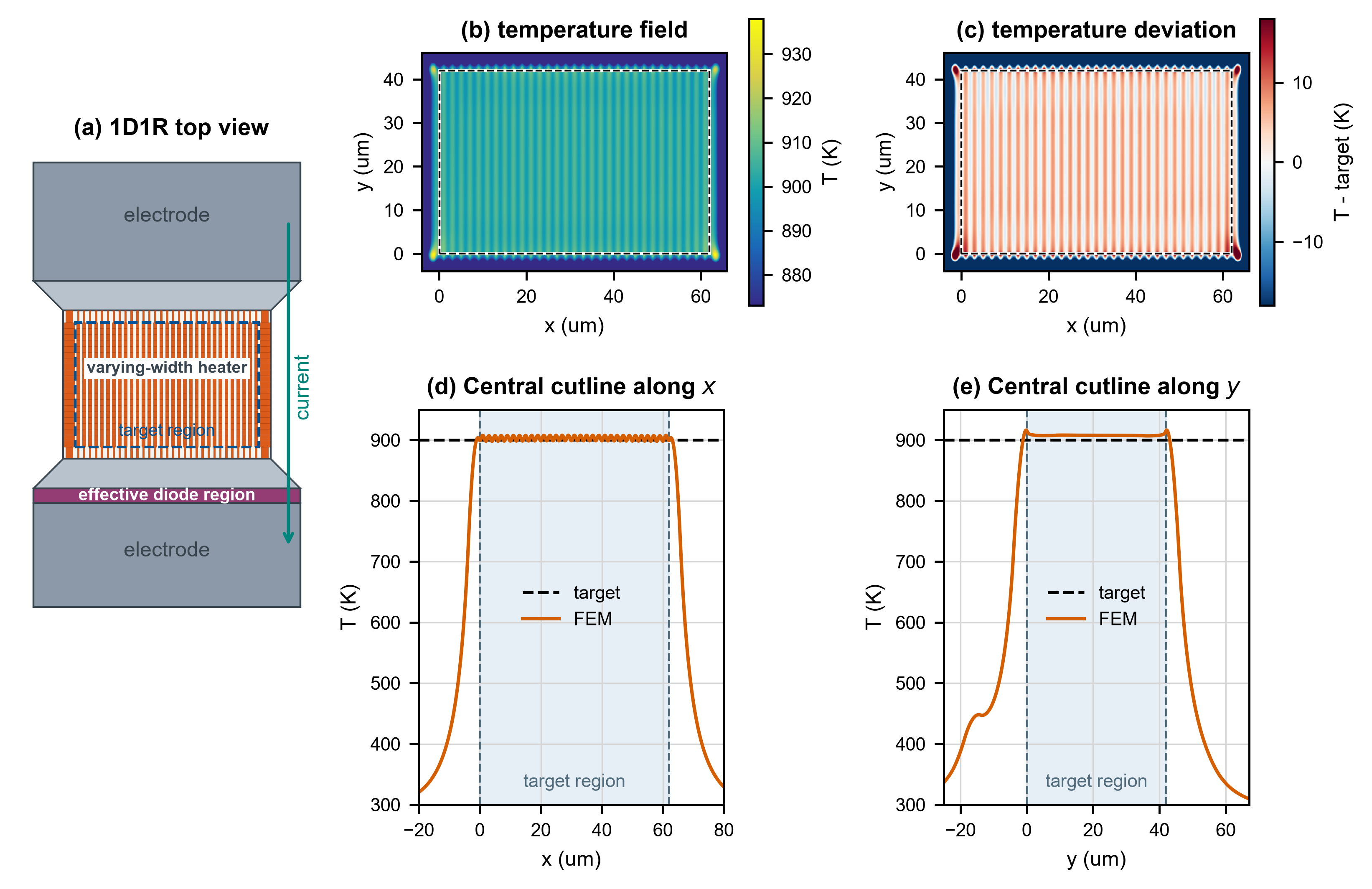}
    \caption{Inverse-designed heater in the 1D1R architecture and its temperature distribution at $t=13~\mu\mathrm{s}$. (a) Top-view geometry showing the relative positions of the electrodes, varying-width heater, target region, and effective diode-associated strip. (b,c) Temperature and temperature deviation from the prescribed $900~\mathrm{K}$ target, evaluated at the mid-plane of the silicon device layer. (d,e) Temperature profiles along the $x$ and $y$ directions through the center of the target region, respectively. The dashed rectangles in (b,c) and shaded intervals in (d,e) indicate the target region. The extended $y$-directed profile in (e) reveals the diode-associated temperature shoulder outside the target region, while the temperature within the target region remains close to the prescribed value. The remaining periodic variation along $x$ arises from the temperature difference between the centers of the conducting filaments and the gaps between adjacent filaments.}
    \label{fig:foundry_2um}
\end{figure*}

The series-connected 1D1R configuration follows the selector-integrated heater architecture developed for a two-dimensional pixel-addressable photonic platform\cite{popescu2026twoDimensionalAddressableMetasurface}. Figure~\ref{fig:foundry_2um}(a) shows the modeled top-view geometry. The $70~\mu\mathrm{m}\times50~\mu\mathrm{m}$ heater is connected through tapered silicon leads to two electrodes, with a $90~\mu\mathrm{m}\times5~\mu\mathrm{m}$ diode-associated silicon strip placed between the lower taper and lower electrode. In the electrothermal model, this strip is assigned an effective conductivity equal to one tenth of the conductivity used for the doped heater region. This reduced-order representation captures the series resistance, localized Joule heating, and one-sided thermal disturbance relevant to the heater design; it does not model the full carrier-transport or current--voltage response of the PIN junction.

Because the diode-associated strip is located on only one side of the heater, its heat generation breaks the symmetry of the temperature field along the $y$ direction. The virtual heat-source grouping is modified accordingly. The center and paired side regions retain their shared variables, whereas the top edge, bottom edge, top corners, and bottom corners are assigned separate powers, as shown in Fig.~\ref{fig:heater_layout}(c). This six-group representation allows the inverse design to reduce heating near the diode and redistribute power elsewhere in the heater. The nonlinear correction step further accounts for changes in the diode-associated heating as the filament widths and total current are updated.

A $30~\mathrm{V}$ pulse is applied to the complete series-connected structure, and the temperature field is evaluated at $t=13~\mu\mathrm{s}$. The initial linewidth-mapped heater exhibits a distorted temperature profile because the linear response model does not fully capture temperature-dependent material properties, current redistribution, and the additional heating associated with the diode configuration. Over eight nonlinear correction iterations, the group powers and filament widths are updated to compensate for these effects. The correction history in Fig.~S2 of the Supplementary Material shows how the temperature distortion near the diode is reduced while the rest of the target region is brought toward the prescribed temperature.

Figure \ref{fig:foundry_2um} shows the final temperature field at the mid-plane of the silicon device layer. The extended $y$-directed profile in Fig.~\ref{fig:foundry_2um}(e) exhibits a distinct temperature shoulder on the diode side outside the target region, directly revealing the localized diode-associated heating. Despite this one-sided thermal input, the inverse-designed heater maintains a temperature close to the prescribed $900~\mathrm{K}$ across most of the target region. Within the target region, the $\mathrm{RMSE}_{\mathrm{field}}$ is $4.8~\mathrm{K}$, and the maximum positive and negative deviations are $+18.0~\mathrm{K}$ and $-6.9~\mathrm{K}$, respectively.

The remaining periodic temperature variation originates from the discrete filament geometry. Along the $x$ direction, the temperature reaches a local maximum near the center of each conducting filament and decreases in the gap between adjacent filaments. The optimization uses the mean of the two transverse-boundary temperatures in each cell, near these local minima. In contrast, within the target region, the central cutline along $y$ follows the center of a filament and therefore passes through the local temperature maxima. This difference explains why the $y$-directed profile in Fig. \ref{fig:foundry_2um}(e) lies slightly above the prescribed temperature. Averaged over all target-region cells, the filament-center temperature exceeds the mean of the two adjacent boundary temperatures by $9.1~\mathrm{K}$, compared with $2.5~\mathrm{K}$ in the $1~\mu\mathrm{m}$-cell heater. This diagnostic quantifies the larger within-cell temperature variation produced by the coarser heater pitch.

The final layout has a minimum filament-segment width of $614.3~\mathrm{nm}$ and a minimum edge-to-edge spacing of $756.6~\mathrm{nm}$. These dimensions are substantially larger than the corresponding values of $280.6~\mathrm{nm}$ and $502.3~\mathrm{nm}$ in the $1~\mu\mathrm{m}$-cell design, demonstrating the improvement in fabrication tolerance obtained by increasing the cell size. The trade-off is a larger temperature variation within each cell, as summarized in Table \ref{tab:uniform_design_comparison}.

To verify that the periodic temperature variation is physical rather than a numerical artifact, the maximum FEM element size in the active silicon is reduced from $400~\mathrm{nm}$ to $300~\mathrm{nm}$ while keeping the geometry, material properties, and electrical excitation unchanged. The refined calculation changes the temperature field by only $0.2~\mathrm{K}$ root mean square within the target region. The temperature range on the central cutline along $x$ changes from $10.0~\mathrm{K}$ to $9.9~\mathrm{K}$. The close agreement confirms that the variation between the filament centers and the intervening gaps is insensitive to further mesh refinement, as shown in Fig.~S3 and Table~S3.

\begin{table*}[t]
    \centering
    \caption{Comparison of the two inverse-designed heaters targeting a uniform temperature distribution at $13~\mu\mathrm{s}$. $\mathrm{RMSE}_{\mathrm{field}}$ and maximum positive and negative errors are calculated relative to $900~\mathrm{K}$ from the two-dimensional target-region temperature fields. The center-to-boundary temperature difference instead quantifies within-cell variation: the cell-center temperature minus the mean of its two transverse-boundary temperatures, averaged over target-region cells. Minimum filament widths and edge-to-edge spacings refer to the complete heater layouts, including the peripheral compensation cells.}    \label{tab:uniform_design_comparison}
    \small
    \begin{tabular}{lcc}
        \hline
        Metric
        & $1~\mu\mathrm{m}$-cell heater
        & $2~\mu\mathrm{m}$-cell heater in 1D1R architecture \\
        \hline
        Target-region dimensions
        & $42~\mu\mathrm{m}\times42~\mu\mathrm{m}$
        & $62~\mu\mathrm{m}\times42~\mu\mathrm{m}$ \\
        Target-relative $\mathrm{RMSE}_{\mathrm{field}}$
        & $3.3~\mathrm{K}$
        & $4.8~\mathrm{K}$ \\
        Maximum positive/negative error
        & $+8.1/-4.4~\mathrm{K}$
        & $+18.0/-6.9~\mathrm{K}$ \\
        \shortstack[l]{Mean center-to-boundary\\temperature difference}
        & $2.5~\mathrm{K}$
        & $9.1~\mathrm{K}$ \\
        Minimum filament width
        & $280.6~\mathrm{nm}$
        & $614.3~\mathrm{nm}$ \\
        Minimum edge-to-edge spacing
        & $502.3~\mathrm{nm}$
        & $756.6~\mathrm{nm}$ \\
        \hline
    \end{tabular}
    \label{tab:uniform_heater_comparison}
\end{table*}

\FloatBarrier
\subsection{Heater for Producing a Linear Temperature Gradient}

The third example demonstrates the generation of a prescribed nonuniform temperature distribution. The heater retains the $70~\mu\mathrm{m}\times50~\mu\mathrm{m}$ dimensions and $33\times23$ design-cell arrangement used in Sec. \ref{sec:rectangular_heater}. However, the series-connected diode and its associated material regions are removed so that the heater remains symmetric along the $y$ direction. The target region measures $62~\mu\mathrm{m}\times42~\mu\mathrm{m}$ and contains a central $31\times21$ array of $2~\mu\mathrm{m}\times2~\mu\mathrm{m}$ cells, surrounded by $4~\mu\mathrm{m}$-wide peripheral compensation cells.

The prescribed temperature increases linearly along the $x$ direction and remains constant along the $y$ direction:

\begin{equation}
    T_{\mathrm{tar}}(x)
    =
    900~\mathrm{K}
    +
    \left(
        2~\mathrm{K}\,\mu\mathrm{m}^{-1}
    \right)
    \left(
        x-x_{\mathrm{c}}
    \right),
    \label{eq:linear_temperature_target}
\end{equation}

\noindent where $x_{\mathrm{c}}$ denotes the center of the target region. The target temperature therefore increases from $838~\mathrm{K}$ at one boundary of the $62~\mu\mathrm{m}$-wide target region to $962~\mathrm{K}$ at the opposite boundary, with a value of 900 K at its center.

The cell grouping is modified to provide the spatial degrees of freedom needed to reproduce this gradient. Each of the 33 columns, including the two peripheral compensation columns, has one independently optimized power for its interior cells. A second power variable is shared by the top and bottom cells of the same column. The latter allows the design to compensate for enhanced heat loss at the upper and lower boundaries while preserving symmetry along $y$. The complete inverse problem therefore contains 66 power variables. A second-difference regularization term is applied along $x$ to suppress abrupt changes between neighboring columns while retaining the prescribed linear variation.

Converting the response-matrix solution into filament widths changes the realized electrothermal response: temperature-dependent material properties and current redistribution cause the nonlinear temperature field to depart from the linear prediction. Nonlinear correction therefore remains necessary for the gradient target. Over nine correction iterations, the spatial power distribution and filament widths are adjusted to bring the temperature profile toward the prescribed gradient.

\begin{figure*}[htbp]
    \centering
    \includegraphics[width=\linewidth]{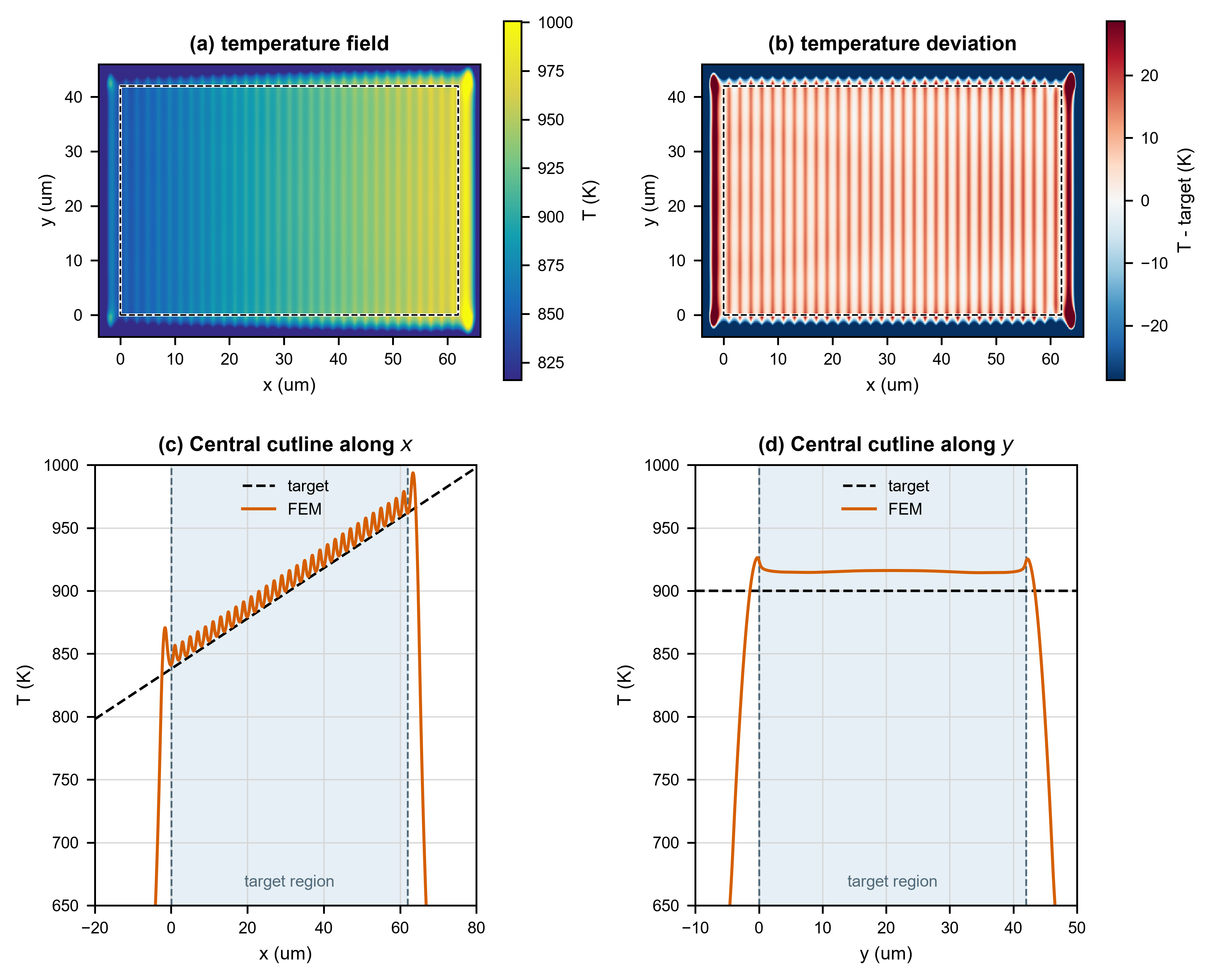}
    \caption{Temperature distribution produced by a heater designed to generate a linear temperature gradient at $t=13~\mu\mathrm{s}$, evaluated at the mid-plane of the silicon device layer. (a) Simulated temperature distribution. (b) Difference between the simulated and prescribed temperatures. (c,d) Temperature profiles along the $x$ and $y$ directions through the center of the target region, parallel and perpendicular to the prescribed gradient, respectively. The dashed rectangles in (a,b) and shaded intervals in (c,d) indicate the target region. Small repeating variations along $x$ arise from the temperature difference between the centers of the conducting filaments and the gaps between adjacent filaments.}
    \label{fig:slope_profile_result}
\end{figure*}

Figure \ref{fig:slope_profile_result} shows the final electrothermal simulation result at $t=13~\mu\mathrm{s}$. The temperature increases approximately linearly along $x$ and remains nearly constant along $y$, demonstrating that the filament widths can be designed to produce a prescribed temperature gradient rather than only a uniform temperature distribution. Within the target region, $\mathrm{RMSE}_{\mathrm{field}}$ is $9.1~\mathrm{K}$, and the maximum positive and negative deviations from the prescribed linear target are $+28.7~\mathrm{K}$ and $-8.5~\mathrm{K}$, respectively. The central cutline along $x$ follows the prescribed slope, with the small temperature oscillations of the individual filaments superimposed on it. The nonlinear-correction history and convergence metrics are reported in Sec.~S2.2 of the Supplementary Material.

Small repeating temperature variations remain along $x$ because heating is localized within the discrete doped silicon filaments. The temperature reaches a local maximum near the center of each filament and decreases in the gap between adjacent filaments. The optimization uses the mean of the two transverse-boundary temperatures in each cell, near these local minima. In contrast, the central cutline along $y$ follows the center of one filament and therefore passes through the local temperature maxima. This sampling difference explains the small offset between the simulated $y$-directed profile and the prescribed temperature in Fig. \ref{fig:slope_profile_result}(d). The results demonstrate that the inverse-design method can generate a spatially varying temperature field while maintaining a continuous heater connected through only one pair of electrodes. The final filament-width maps for all three designs are shown in Fig.~S1, and their minimum filament widths and edge-to-edge spacings are summarized in Table~S2.

\FloatBarrier
\section{Discussion}

A central consideration in the inverse problem is solution stability. Thermal diffusion causes neighboring heat-source cells to produce strongly correlated temperature responses. Consequently, a pixel-by-pixel inversion, in which every cell is assigned an independent power density, can be ill-conditioned and admit multiple power distributions that produce nearly indistinguishable temperature fields. Such solutions may contain large, rapidly alternating power-density variations between adjacent cells. These variations provide little improvement in the resulting temperature distribution because they are smoothed by thermal diffusion. They are also sensitive to small modeling or fabrication errors and may require abrupt changes in filament width that are difficult to implement. Grouping cells with equivalent physical roles removes these unnecessary degrees of freedom and restricts the solution to smoother and more robust power distributions. In this sense, grouping provides a physically informed form of regularization. This approach differs from the Tikhonov regularization used by Tan et al. for an independently controlled heater array\cite{tan2017inverseHeatTransferMicroHeaterArray}, as the present grouping is based on the geometry, symmetry, and electrical connectivity of a single heater.

The appropriate grouping depends on the target temperature distribution, heater geometry, and thermal boundary conditions. In the present work, the groups are selected with human intervention based on the expected symmetries and heat-loss pathways. The uniform-temperature heater in Sec. \ref{subsec:1um} uses four groups because both the heater and its thermal boundary conditions are symmetric along $x$ and $y$. In the 1D1R configuration, the diode introduces asymmetric heating along $y$, as reflected by the diode-side temperature shoulder in Fig.~\ref{fig:foundry_2um}(e). Separate power variables are therefore used for the upper and lower edges and corners, increasing the number of groups to six. The linear-gradient design requires independent spatial variation along $x$ and therefore uses two variables for each column, giving 66 groups. Thus, a grouping selected for one geometry or boundary condition cannot necessarily be transferred to another. In future implementations, group selection could be incorporated into a fully automated, end-to-end inverse-design process. For example, cells could be grouped algorithmically according to the similarity of their transient response-matrix columns, while simultaneously accounting for the target field, heater connectivity, and fabrication constraints.

Fabrication-induced linewidth changes alter the heater resistance, total heating power, and mean temperature. We examine a systematic width error by decreasing or increasing every filament-segment width in the final $2~\mu\mathrm{m}$-cell heater by $50~\mathrm{nm}$ without reoptimizing the geometry (Fig.~S4 and Table~S4 in the Supplementary Material). For each modified layout, the pulse-voltage amplitude is adjusted to restore the target-region mean temperature of the nominal layout. The required voltages are $30.86~\mathrm{V}$ and $29.19~\mathrm{V}$ for the $-50~\mathrm{nm}$ and $+50~\mathrm{nm}$ layouts, respectively, compared with $30~\mathrm{V}$ for the nominal layout. Voltage compensation restores the mean temperature while retaining a largely uniform temperature distribution. The target-region spatial standard deviation after compensation is $6.18~\mathrm{K}$ for the $-50~\mathrm{nm}$ layout and $5.78~\mathrm{K}$ for the $+50~\mathrm{nm}$ layout, compared with $3.49~\mathrm{K}$ for the nominal layout. On the central cutline along $y$, the temperature range within the target region is $27.46~\mathrm{K}$ for the $-50~\mathrm{nm}$ layout and $14.82~\mathrm{K}$ for the $+50~\mathrm{nm}$ layout, compared with $8.69~\mathrm{K}$ for the nominal layout. These residual changes arise because the narrow boundary-compensation segments undergo the largest fractional resistance change, which cannot be fully offset by a single voltage adjustment.

The linewidth-bias calculation isolates one systematic fabrication error. Other nonidealities, including dopant diffusion, implantation-dose variations, line-edge roughness, layer-thickness variations, contact resistance, and uncertainty in temperature-dependent material properties, may also modify the current and temperature distributions. These effects can be incorporated into future models through measured process distributions, worst-case constraints, or optimization over an ensemble of perturbed geometries\cite{popescu2026lossless}. Such extensions would enable heater designs that are optimized not only for nominal performance but also for robustness to realistic fabrication variations. Experimental validation will ultimately be required to quantify these effects and establish process-specific design margins.

\section{Conclusion}

We developed an inverse-design framework for tailoring transient temperature distributions in resistive microheaters. The method combines grouped heat-source optimization, fabrication-constrained mapping of the optimized power distribution to doped-silicon filament widths, and nonlinear electrothermal correction. We demonstrated uniform heating, compensation of diode-induced thermal asymmetry in a 1D1R architecture, and generation of a prescribed linear temperature gradient. These results establish a general, fabrication-compatible approach for designing resistive microheaters with prescribed transient temperature distributions.

\section*{Supplementary Material}
See the supplementary material for additional finite-element model details, optimized linewidth maps, nonlinear-correction histories, mesh-refinement results, and a systematic filament-width-error analysis.

\begin{acknowledgments}
The authors acknowledge funding support provided by the Air Force Office of Scientific Research (AFOSR) under award FA9550-22-1-0532 and the National Science Foundation under award 2328839. This work was also supported by National Research Foundation of Korea (NRF) grants funded by the Korean government (Ministry of Science and ICT, MSIT) in 2026 (Grant Nos. RS-2025-00515651 and RS-2025-02213804).

OpenAI Codex (primarily GPT-5.5; \url{https://openai.com/codex/}) assisted code development and debugging, and language editing of the manuscript. The authors are responsible for the final content.
\end{acknowledgments}

\section*{Author Declarations}

\subsection*{Conflict of Interest}
The authors have no conflicts to disclose.

\subsection*{Author Contributions}
\textbf{Muzhang Huang:} Conceptualization; Methodology; Software; Investigation; Formal analysis; Visualization; Writing--original draft; Writing--review \& editing.

\textbf{Junhe Zhang:} Methodology; Resources; Project administration; Writing--review \& editing.

\textbf{Khoi Phuong Dao:} Software; Resources; Writing--review \& editing.

\textbf{Juejun Hu:} Supervision; Funding acquisition; Writing--review \& editing.

\textbf{Hyun Jung Kim:} Supervision; Funding acquisition; Writing--review \& editing.

\section*{Data Availability}
Source code for the inverse-design simulation workflow is publicly available under the BSD 3-Clause License at \url{https://github.com/muzhanghuang-jpg/transient-inverse-thermal-design-code}. The data that support the findings of this study are available from the corresponding author upon reasonable request.

\bibliographystyle{aipnum4-2}
\bibliography{references}

\end{document}